\documentclass[prd,aps,graphics]{revtex4}
\usepackage{epsfig}
\newcommand{\dd}{{\rm d}}
\newcommand{\bk}{{\bf k}}
\newcommand{\bx}{{\bf x}}
\newcommand{\by}{{\bf y}}

\newcommand{\chib}{{\bar \chi}}

\newcommand{\chis}{\chi_{_{\rm S}}}
\newcommand{\dchis}{\delta\chi_{_{\rm S}}}
\newcommand{\chisO}{\chi_{_{\rm S}}^{(0)}}
\newcommand{\dchisO}{\delta\chi_{_{\rm S}}^{(0)}}
\newcommand{\chisOb}{{\overline{\chi_{_{\rm S}}^{(0)}}}}
\newcommand{\chih}{\chi_{_{\rm H}}}
\newcommand{\chir}{\chi_{_{\rm R}}}
\newcommand{\rs}{R_{_{\rm S}}}
\newcommand{\rh}{R_{_{\rm H}}}
\newcommand{\ks}{k_{_{\rm S}}}
\newcommand{\kh}{k_{_{\rm H}}}
\newcommand{\rphys}{R^{(\rm phys)}}
\newcommand{\what}{{\widehat W}}
\newcommand{\mg}{\left<}
\newcommand{\md}{\right>}
\newcommand{\mC}{{\widetilde W}}
\newcommand{\mP}{{\cal P}}
\newcommand{\mW}{{\cal W}}

\begin{document}
\title{Finite volume effects for non-Gaussian multi-field inflationary models}

 \author{Francis Bernardeau}
 \email{fbernard@spht.saclay.cea.fr}
 \affiliation{Service de Physique Th{\'e}orique, CEA/DSM/SPhT,
              Unit{\'e} de recherche associ{\'e}e au CNRS,
              CEA/Saclay 91191 Gif-sur-Yvette c{\'e}dex}

 \author{Jean-Philippe Uzan}
 \email{uzan@th.u-psud.fr}
 \affiliation{Laboratoire de Physique Th{\'e}orique, CNRS--UMR 8627,
              B{\^a}t. 210, Universit{\'e} Paris XI,
              F-91405 Orsay Cedex, France,\\
              Institut d'Astrophysique de Paris, GReCO,
              CNRS-FRE 2435, 98 bis, Bd Arago, 75014 Paris, France.}

\date{\today}
%%%%%%%%%%%%%%%%%%%%%%%%%%%%%%%%%%%%%%%%%%
\begin{abstract}
Models of multi-field inflation exhibiting primordial
non-Gaussianity have recently been introduced. This is the case in
particular if the fluctuations of a light field scalar field,
transverse to the inflaton direction, with quartic coupling can be
transferred to the metric fluctuations. So far in those
calculations only the ensemble statistical properties have been
considered. We explore here how finite volume effects could affect
those properties. We show that the expected non-Gaussian
properties survive at a similar level when the finite volume
effects are taken into account and also find that they can skew
the metric distribution even though the ensemble distribution is
symmetric.
\end{abstract}
\pacs{{\bf PACS numbers:} 98.80.-q, 04.20.-q, 02.040.Pc}
\vskip2pc
%%%%%%%%%%%%%%%%%%%%%%%%%%%%%%%%%%%%%%%%%%
\maketitle

%%%%%%%%%%%%%%%%%%%%%%%%%%%%%%%%%%%%%%%%%%
\section{Introduction}\label{sec_intro}

The observations of Cosmic Microwave Background anisotropies and
of the large-scale structure of the universe offer a window to the
physics of the inflaton. For instance the detailed measurement of
the shape of the power spectrum can give constraints on the shape
of the inflaton potential. On the other hand the detection of
Non-Gaussian metric fluctuations could signal the existence of
effective couplings between the inflaton and other fields. There
have been a series of observational progresses in view of
constraining the deviation from Gaussianity using CMB data on
large scales ($\theta\sim10^o$)~\cite{large}, intermediate
($\theta\sim1^o$)~\cite{intermediate} and small scales
($\theta\sim10'$)~\cite{small}, as well as large scales
structures~\cite{lss}. The recent CMB data by WMAP were also used
by Komatsu {\em et al.}~\cite{komatsu} that concluded that the CMB
spectrum was compatible with Gaussianity on the basis of an
analysis of the bispectrum. Similar analysis using Minkowski
functionals~\cite{colley} and 3-point correlation
function~\cite{gaztanaga} reached the same conclusions. More
recently, an analysis of the WMAP data using measurements of the
genus and its statistics~\cite{park} conluded that the Gaussianity
of the CMB field was ruled out at a 99\% level. As stressed after,
non-Gaussianity of primordial origin may still have escaped
detection and further investigations are needed.

From a theoretical point of view, Gaussianity is a generic
prediction of slow-roll single field
inflation~\cite{maldacena,bu1}. A series of models including
feature in the inflationary potential~\cite{features}, existence
of seeds such as $\chi^2$~\cite{chi2}, axion~\cite{axion} or
topological defects~\cite{topdef}, curvaton
scenario~\cite{curvaton}, varying inflaton decay rate~\cite{decay}
have been shown to be able to generate some primordial
non-Gaussianity.

In Ref.~\cite{bu1}, we proposed a general mechanism to produce
non-Gaussianity in the adiabatic mode and explicit realizations of
such models were presented in Ref.~\cite{bu2}. The mechanism is
based on the generation of non-Gaussian isocurvature fluctuations
which are then transferred to the adiabatic modes through a bend
in the classical inflaton trajectory. Natural realizations were
shown to involve quartic self-interaction terms. The statistical
properties of the resulting metric fluctuations were then shown to
be the superposition of a Gaussian and a non-Gaussian contribution
of the same variance, the relative weight of the two contributions
being related to the total bending of the trajectory in field
space. The non-Gaussian probability distribution function (PDF)
was also computed and shown to be described by a single new
parameter so that generically only two new parameters suffice in
describing this class of models.

In order to infer constraints from e.g. CMB data on these kind of
models predicting non-Gaussianity one needs at least to go through
2 steps.
\begin{enumerate}
\item One first needs a precise prediction of the statistical
properties of the curvature in order to construct estimator
adapted to the detection of this kind of non-Gaussianity. For
instance the analysis Komatsu {\em et al}.~\cite{komatsu} uses the
bispectrum and Minkowski functional and constrains a $\chi^2$
deviation from Gaussianity. The gravitational potential was
parametrized as $\Phi=\Phi_{\rm L} + f_{\rm NL}(\Phi^2_{\rm L} -
\left<\Phi^2_{\rm L}\right>)$ where $\Phi_{\rm L}$ is the Gaussian
linear perturbation of zero mean and it was concluded that
$-58<f_{\rm NL}<134$ at 95\% CL. This constrains only a very
peculiar type of non-Gaussianity and does not apply e.g. to the
models of Refs.~\cite{bu1,bu2}.
\item Even if the form of the PDF is predicted theoretically, one
needs to investigate what is measured and how the measurements are
related to the theoretical predictions.
\end{enumerate}

The goal of this article is to investigate the observational
implications of the theoretical predictions of Ref.~\cite{bu1}. It
requires further investigations in particular on the effect of
finite volumes of the survey while the detailed comparison of our
model to existing dataset, beyond the scope of this work, is still
left for further studies. In particular, we want to show that
these finite volume effects do not suppress the predicted
non-Gaussianity.

Section~\ref{sec_pb} formulates the problem, focusing on the
quantities that can actually be observed. In
Section~\ref{sec_mod}, we recall basics of the models we consider
in this paper and give simple consequences. We then develop in
Section~\ref{sec_langevin} some arguments to understand the effect
of the mean value of the non-Gaussian component on the size of the
survey. Section~\ref{sec_finitevol} is devoted to the computation
of the lowest order measured cumulants. In particular, we will
show that there appears a non vanishing skewness that is directly
induced by the finite volume effects as stressed in our concluding
remarks, Section~\ref{sec_concl}.

%%%%%%%%%%%%%%%%%%%%%%%%%%%%%%%%%%%%%%%%%%%%%%%%%%%
\section{An overview of finite volume effects}\label{sec_pb}

Basically what we want to measure are the statistical properties
of a scalar field $\chi$ of zero mean
$$
\left<\chi\right>=0,
$$
where and from now on angular brackets,$\left<\,\,\right>$, refer
specifically to ensemble averages. Such a scalar field will be in
the following identified with the gravitational potential but the
actual observations may be complex linear transforms of that field
such as the temperature anisotropies or polarization of the CMB or
the large-scale cosmic density fluctuations. The general questions
raised by precision measurements of statistical properties of
cosmic fields can be very intricate. We refer for instance
to\cite{revue} Section~VI for extensive developments regarding
such issues. We will restrict our discussion to the overall
aspects of them.

In practice while performing such measurements, two scales enter
the problem, the scale $\rs$ at which $\chi$ is measured (e.g.
somehow smoothed) and the size $\rh$ of the survey. The smoothed
field $\chis$ can be measured at different locations in the
survey. The idea is that its spatial fluctuations should provide
us with hints of the actual statistical properties of the field.
Ensemble averages are however inaccessible as such. The values of
$\chis$ we have access to are in finite number and are all
inherited from the a single stochastic process, the one of our
universe. One can however get insights into the stochastic
properties of $\chi$ from geometrical averages. In the following,
we denote such a geometrical average $\overline{A}$ and its
connected part $\overline{A}^{\,_c}$. For instance one can have
access to quantities such as
$$
X_n \equiv \overline{({\chis}-\chib)^n\,}^{\,_{c}} =
\overline{\delta{\chis}^n\,}^{\,_{c}}.
$$
The $X_n$ are themselves stochastic quantities that depend on the
stochastic field $\chi$. To be more specific we have
\begin{equation}
 \chib=\int \chi(\by,t){\cal W}_{\rh}(\by)\dd^3\by.
\end{equation}
and
\begin{equation}\label{defchib}
 \chis(\bx)=\int \chi(\by,t){\cal W}_{\rs}(\bx-\by)\dd^3\by
\end{equation}
where ${\cal W}_{R}$ is a window function of volume unity and for
simplicity assumed to be spherically symmetric, e.g. for a top hat
filtering we have $\mW_R=\Theta(|\bx-\by|-R)/V$ where $\Theta$ is
the heavyside function and $V$ the volume of the ball or
$\mW_R=\exp(-|\bx-\by|^2/(2R)^2)/(\sqrt{2\pi}R)^3$ for a Gaussian
window function.

Clearly quantities such as $\left<\chi^2\right>$ cannot be
estimated because they contain contributions from modes of
wavelength larger than the size of the survey and that cannot be
observed. Only modes with wavelength in a given range are
measurable. This is in particular the case for a Gaussian field.
In this case one can show that if the two scales $\rs$ and $\rh$
are in large enough ratio, $X_n$ can be good estimates of
$\left<\left(\chis-\chib\right)^n\right>$, e.g. the expected
dispersion of the measured values $X_n$ decreases with $\rh/\rs$
to a power that depends on the power spectrum shape. In our case
however not only the super-Hubble modes cannot be measured but the
modes that can be observed are correlated and also correlated to
the super-Hubble modes.

In the following, we will explore the consequences of both the
fact that only modes in specific wavelength are measurable and
that they are correlated. Ideally, cosmological models should be
able to predict the PDFs of the measured cumulants,
$$
{\cal P}(X_n),
$$
but such predictions are obviously difficult to do in general. We
will see in the following how we can estimate these distributions
for some families of multi-field inflationary models.

%%%%%%%%%%%%%%%%%%%%%%%%%%%%%%%%%%%%%%%%%%%%%%%%%%%
\section{Models}\label{sec_mod}

\subsection{Self-interacting scalar fields in de Sitter space}\label{subsec1}

Before we start to investigate the statistics of the observed
quantities let us recall the basics of the models we have in mind.
In the models~\cite{bu1,bu2}, the non-Gaussianities are first
generated by an auxiliary field, $\chi$, self-interacting in a
potential, typically quartic,
\begin{equation}\label{Vquartic}
  V(\chi)=\frac{\lambda}{4!}\chi^4.
\end{equation}
This field is assumed to be a test field so that it does not
affect the dynamics of inflation driven by an other scalar field,
$\phi$. Assuming an almost de Sitter inflation and neglecting the
gravitational back-reaction on the evolution of the universe, its
evolution will be dictated by
\begin{equation}\label{KGchi}
 \ddot\chi+3H\dot\chi-\frac{1}{a^2}\Delta\chi=-\frac{\lambda}{3!}\chi^3
\end{equation}
where $H=\dot a/a$ is the Hubble constant and $\Delta$ the
comoving Laplacian. The evolution of the scale factor is given by
\begin{equation}\label{bgd}
 a(t)=a_0\hbox{e}^{Ht},\qquad
 \eta=-\frac{\hbox{e}^{-Ht}}{H},\quad
 a(\eta)=-\frac{1}{H\eta}
\end{equation}
where $t$ is the cosmic time and $\eta$, which is negative, the
conformal time.

For a free massless scalar field ($\lambda=0$), the Klein-Gordon
equation (\ref{KGchi}) in the de Sitter background takes the form
\begin{equation}\label{KGchi2}
 v''+\left(k^2-\frac{2}{\eta^2}\right)v=0
\end{equation}
where we have introduced $v\equiv \chi/a$. The general of this
equation is
\begin{equation}\label{chisol}
 \chi_k(t)=-\frac{H\eta}{\sqrt{2k}}\left(1+\frac{1}{ik\eta}\right)\,
 \hbox{e}^{-ik\eta},
\end{equation}
once the quantum field, $\chi$, is decomposed in Fourier modes
as\footnote{Note that the convention of Eq.~(\ref{fourier})
differs by a factor $(2\pi)^{3/2}$ compared to the one of
Ref.~\cite{bbu}. It implies that Eq.~(\ref{chi4pt}) differs by a
factor $(2\pi)^6$ compared to its original form.}
\begin{equation}\label{fourier}
 \chi(\bx,t)=\int\frac{\dd^3\bk}{(2\pi)^{3/2}}\left[
 \chi_k(t)\hbox{e}^{i\bk.\bx}\hat b_\bk + \hbox{h.c.}
 \right],
\end{equation}
where $[\hat b_\bk,\hat b_{\bk'}^\dag]=(2\pi)^3\delta(\bk-\bk')$.

It implies that the field $\chi$ has a correlator given by
\begin{equation}\label{PdeK}
\left<\chi(\bk)\chi^*(\bk')\right>= (2\pi)^3P(k)\delta(\bk-\bk')
\end{equation}
where $P(k)$ is the power spectrum, which is equal to
$P(k)=H^2/2k^3$ on super-Hubble scales for a free scalar field
living in de Sitter space. This is the so-called
Harrison-Zel'dovich spectrum.

Moreover, the self interaction term of $\chi$ induces non-zero
high order correlation functions. For a quartic potential the odd
order correlation functions vanish; the even order ones can be
computed from a perturbation theory approach at tree order. For
instance, the four-point function in the super-Hubble limit reads
\begin{equation}\label{chi4pt}
\left<\chi_{\bk_1}\ldots\chi_{\bk_4}\right>_c
=-\frac{\lambda}{3}(2\pi)^6\frac{\log \left(\eta\sum_i k_i \right)}
{H^2}\,\delta\left(\sum_i\bk_i\right)\,P(k_1)P(k_2)P(k_3)+
 \,{\rm perm.},
\end{equation}
as explicitly shown in Ref.~\cite{bbu}.

%%------------------------------------------------------------
\subsection{Second and fourth moment}\label{subsec_low}

When one wants to relate the expectation values of observable
quantities such as the $X_n$ to the statistical properties of the
field, filtering effects should properly be taken into account.
The observable quantities, $\delta\chis$, can be written in terms
of the Fourier modes, $\chi_{\bk}$, as
\begin{equation}
\delta\chis=\int\frac{\dd^3\bk}{(2\pi)^{3/2}}\hbox{e}^{i\bk.\bx}\chi(\bk)\,\widetilde
W(k)
\end{equation}
with
\begin{equation}\label{wtilde}
\widetilde W(k) \equiv \widehat W(k\rs) - \widehat W(k\rh).
\end{equation}
It follows immediately that
\begin{equation}
 \mg\delta\chis^2\md = 4\pi \int k^2 P(k) \widetilde W^2(k)\,
 \dd k.
\end{equation}
Such an expression can be easily computed in case of a
Harrison-Zel'dovich spectrum and a $k$-space top-hat window
function. Defining $\ks$ and $\kh$ respectively as $1/\rs$ and
$1/\rh$, the second moment of $\delta\chis$ reduces to
\begin{equation}
 \mg\delta\chis^2\md = 2\pi H^2
 \ln\left(\frac{\ks}{\kh}\right)\equiv\sigma_\delta^2.
\end{equation}

It follows from this calculation that the expectation value of the
observable quantity $X_2$ is given by
\begin{equation}
 \left<X_2\right> = \mg\delta\chis^2\md = \sigma_\delta^2.
\end{equation}
Its (cosmic) variance, $\left<X_2^2\right>_c^{1/2}$, can similarly
be computed and it scales like $\rs/\rh$ in case of a Gaussian
field~\footnote{The cosmic variance for the measured $X_2$ is
expected to be strongly affected by the existence of a
non-vanishing large-scale 4-point function for $\chi$. But this is
beyond the scope of this paper to investigate such effects.}.

Furthermore, the expectation value of $X_4$ is given by
\begin{equation}\label{X4expression}
\left<X_4\right> = \mg\delta\chis^4\md_c
 =\int\frac{\dd^3\bk_1\ldots\dd^3\bk_4}{(2\pi)^6}\widetilde
 W(k_1)\ldots\widetilde
 W(k_4)\,\left<\chi_{\bk_1}\ldots\chi_{\bk_4}\right>_c.
\end{equation}
Expressing the 4-point correlator by mean of Eq.~(\ref{chi4pt})
and assuming a Harrison-Zeldovich spectrum and a top-hat window
function in $k$-space, this expression finally reads
\begin{equation}\label{kurtosisexpression}
\mg\delta\chis^4\md_c=-{4 \lambda\over 3 H^2}
\left({H^2\over2}\right)^3 \int_{\kh}^{\ks}{\dd^3 \bk_1\over
k_1^3} \int_{\kh}^{\ks}{\dd^3 \bk_2\over k_2^3}\
\int_{\kh}^{\ks}{\dd^3 \bk_3\over k_3^3}\ \ \widetilde
W\left(\vert\bk_1+\bk_2+\bk_3\vert\right)
\log\left[\left(k_1+k_2+k_3+\vert\bk_1+\bk_2+\bk_3\vert\right)\eta\right].
\end{equation}
where $\mC\left(\vert\bk_1+\bk_2+\bk_3\vert\right)$, defined in
Eq.~(\ref{wtilde}), simply expresses the condition $\kh <
\vert\bk_1+\bk_2+\bk_3\vert < \ks$ and is inherited for the term
$\widetilde W(k_4)$. When the ratio $\ks/\kh$ is large this
expression can be easily computed after noting that, in this case,
one of the $k_i$ generically dominates over the others (the
contributions to the integral being roughly uniform over the
integration domain when the wave-vector norms are logarithmicallly
spaced). In this limit it is therefore possible to approximate
$\bk_1+\bk_2+\bk_3$ by say $\bk_1$ and $k_1+k_2+k_3+\vert
\bk_1+\bk_2+\bk_3\vert$ by $2\,k_1$. Finally the integral reads,
\begin{equation}\label{kurtosisexpression2}
  \mg\delta\chis^4\md_c\simeq-{4\over 3}\,\lambda\, \log\left(2\eta
\ks^{3/4}\kh^{1/4}\right)\ {\sigma_{\delta}^6\over H^2}
\end{equation}
which corresponds to what has been found in our previous work
provided the number of $e$-folds is identified with
$\log\left(2\eta \,\ks^{3/4}\,\kh^{1/4}\right)$. It means that the
kurtosis of the $\delta\chis$ field is significant if ${\lambda
\log\left(2\eta\, \ks^{3/4}\,\kh^{1/4}\right)}$ approaches unity.
This result demonstrates that the mechanism described in our
previous study survives finite volume effects. We can also note
that observable quantities are insensitive to the behavior of
$P(k)$ in the small $k$ limit.

Finite volume effects have however other consequences due to the
fact that the observable modes are correlated. The physical reason
of these correlations is that they share the same history, e.g.
they have been produced in the same stochastic process. We found
that a random walk approach for the evolution of the local $\chi$
field values gives precious insights on those more subtle finite
volume effects.

%%%%%%%%%%%%%%%%%%%%%%%%%%%%%%%%%%%%%%%%%%%%%%%%%%%
\section{Lessons from a random walk approach}\label{sec_langevin}

From the previous definitions, it is clear that $\chib$ and the
different values of $\chis$ share contributions of super-hubble
modes. Those cannot be observed but they shape the values of
$\chib$ and $\chis$. Actually the value of $\chir$ at a given
scale, whether it is $\rh$ or $\rs$, is dynamically built from the
stacking of modes that successively leave the horizon.

A random walk approach can then be used to describe the stochastic
growth of $\chir$. It will allow two things, to get insights into
the excursion values of $\chib$ and to see how $\chis$ values are
correlated through their common history. In this approach the
field value evolution is described in term of a Langevin eq
during inflation.

Before we go to this equation, let us sort out how the different
scales and evolution equation and related together.

\subsection{Scales intervening in the problem}

Different scales will have to be distinguished in our study.
\begin{itemize}
\item During the inflationary stage, the super-Hubble modes of the
scalar field can be treated as classical. In a de Sitter
spacetime, the physical Hubble radius is constant $\rphys=H^{-1}$
so that the comoving smoothing scale is time dependent,
$R=(aH)^{-1}$. The evolution equation of the classical part will
thus contain a stochastic force simulating the effect of the
quantum noise due to the modes which are crossing the horizon at
each time step to become classical. We thus define a stochastic
field, $\chih(t)$, for which the filtering scale is dependent and
always equal to the horizon size, $\rh^{\rm phys}=a(t)\rh$ with
$\rh^{\rm phys}\sim H^{-1}$ so that
\begin{equation}\label{scale3}
 \chih=\chi_{R=1/aH}.
\end{equation}
The dynamics of $\chih$ will follow a Langevin equation. We
investigate the dynamics and the statistical properties of $\chih$
in Section~\ref{subsec_langevin}.

\item
On the other hand the field values $\chib$ and $\chis$ can be seen
as a time dependent quantity but they correspond to the filtering
of $\chi$ at fixed physical scales. They identify though with
$\chih$ at precisely the time $\eta$ at which the scales $\rh$ and
$\rs$ respectively cross the horizon, e.g.
$\chib(\eta=-1/\rh{}_{_0})=\chih(\eta= -1/\rh{}_{_0})$ and
$\chis(\eta=-1/\rs)=\chih(\eta= -1/\rs)$. After that these
coincidental times the two fields $\chib$ and $\chis$ behave
classically, i.e. follow an inflationary classical Klein-Gordon
equation without stochastic source terms.
\end{itemize}

\begin{figure}
 \centerline{\epsfig{figure=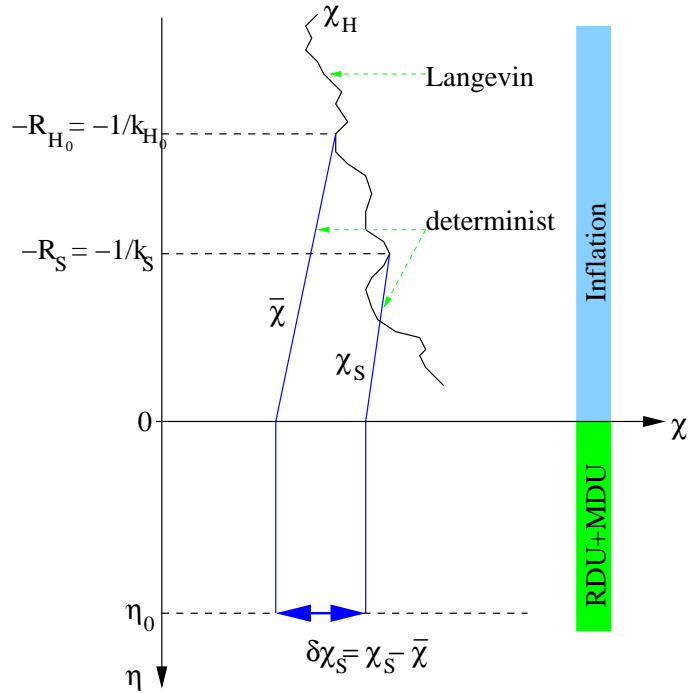,width=9cm}}
 \caption{
The different filtered quantities and scales entering our problem.
$\chih$ follows a stochastic dynamics that can be described by a
Langevin equation; it has a time dependent smoothing scale so that
more and more modes contribute to the filtered field. The field
values $\chib$ and $\chis$ evolve according a classical
Klein-Gordon equation; their smoothing scale is time independent;
they coincide with $\chih$ at horizon crossing times.}
\label{scales}
\end{figure}

To summarize, two of the comoving scales $\rs$ and $\rh{}_{_0}$
are fixed while one, $\rh$ is a time dependent quantity. A sketch
of the different sequences that the field dynamics follow are
shown on Fig.~\ref{scales}.

%%------------------------------------------------------------
\subsection{The late time PDF of $\chih(\eta)$.}\label{subsec_langevin}

We follow a formalism first developed to deal with
self-interacting fields in a de Sitter background~\cite{staro0}
based on the idea that the infrared part of the scalar field may
be treated as a classical spacetime dependent stochastic field
satisfying a Langevin equation~\cite{staro1,staro2} (see
Ref.~\cite{vilenkin} for the case of a massless free field).

Assuming that $\chi$ is slow rolling, the dynamics of $\chih$ can
be obtained by averaging Eq.~(\ref{KGchi}) in which both the
second time derivative and the Laplacian term can be neglected,
since $\chih$ contains only long wavelength modes. Using the
identity
\begin{equation}\label{commutateur}
 \dot\chih{}_\bk=(\chi_\bk)_{R_{\rm H}}\what(k\rh)-k\rh\chi_\bk\what'(k\rh)
\end{equation}
where $\what'(u)=\dd\what(u)/\dd u$ and where we have used
Eq.~(\ref{bgd}) to express the time derivative of $\rh$. It
follows that
\begin{equation}\label{KGsmooth}
 \dot\chih=-\frac{1}{3H}\bar V'(\chih+\delta\chi)+ \xi_{_{\rm Q}}(\bx,t)
\end{equation}
which reduces ($\delta\chi\ll\chih$) to
\begin{equation}\label{KGsmooth2}
 \dot\chih=-\frac{1}{3H}\frac{\dd V(\chih)}{\dd\chih}+ \xi_{_{\rm Q}}(\bx,t).
\end{equation}
The term $\xi_{_{\rm Q}}(\bx,t)$ appears from the commutation
between $({\dot\chi})_{R_{\rm H}}$ and $\dot\chih$ using
Eq.~(\ref{commutateur}). It is a stochastic noise describing the
effects of the small wavelength (quantum) part exiting the horizon
on the classical stochastic part. Using Eq.~(\ref{fourier}), one
obtains its expression as
\begin{equation}\label{noise}
 \xi_{_{\rm Q}}(\bx,t)=-\int\frac{\dd^3\bk}{(2\pi)^{3/2}}(kH\rh)
 \left[\chi_k(t)\what'(k\rh)\hbox{e}^{i\bk.\bx}\hat b_\bk
 + \hbox{h.c.}\right].
\end{equation}
Indeed, $\xi_{_{\rm Q}}$ is quantum noise so that we replace it
heuristically by a Gaussian stochastic noise, $\xi$ with a
correlator that matches the quantum expectation value in the
standard Bunch-Davies vacuum, that is
\begin{equation}\label{noise1}
 \left<\xi(\bx,t)\xi(\bx',t'\right>=
  \left<0\right\vert\xi_{_{\rm Q}}(\bx,t)\xi_{_{\rm
  Q}}(\bx',t')\left\vert0\right>.
\end{equation}
It follows that
\begin{equation}\label{noise2}
 \left<\xi(\bx,t)\xi(\bx',t')\right>=
 \int\frac{k^3\dd k}{4\pi^2}\frac{\sin kr}{r}\frac{\chi_k(t)\chi_k^*(t')}{a(t)a(t')}
 \what'\left(\frac{k}{a(t)H}\right)
 \what'\left(\frac{k}{a(t')H}\right),
\end{equation}
with $r\equiv\vert\bx-\bx'\vert$, which reduces, using the
expression (\ref{chisol}) on small scales, to~\cite{wivi}
\begin{equation}\label{noise3}
 \left<\xi(\bx,t)\xi(\bx',t')\right>=\frac{H^4\eta\eta'}{4\pi^2 r}
 \int\dd k\sin kr (1+ik\eta)(1-ik\eta')\hbox{e}^{ik(\eta'-\eta)}
 \what'\left(-k\eta\right) \what'\left(-k\eta'\right).
\end{equation}
Even though $\chih$ remains a quantum operator, it was replaced by
a stochastic field in a way that, for all observables, the
expectation values of the two fields are in excellent agreement.

In the following and for the sake of simplicity, we consider a
window function reducing to a top-hat in Fourier space so that
$\what'$ reduces to a Dirac distribution. In that case, using the
solution (\ref{chisol}) one obtains
\begin{equation}\label{noise4}
 \left<\xi(\bx,t)\xi(\bx',t')\right>=\frac{H^3}{4\pi^2}
 \frac{\sin a(t)Hr}{a(t)Hr}\delta(t-t').
\end{equation}
The case of more realistic window functions (such as a Gaussian)
was discussed in Ref.~\cite{wivi}.

From a Langevin equation of the form
\begin{equation}\label{chandra1}
 \dot\chih=-\beta V'+ \alpha \xi,\qquad
 \left<\xi(\bx,t)\xi(\bx',t')\right>=\delta(t-t')
\end{equation}
where we assume the two coefficients $\beta$ and $\alpha$ to be
independent of $\chih$, one can deduce~\cite{chandra} a equation,
the Fokker-Planck equation, for the PDF of $\chih$ of the form
\begin{equation}\label{chandra2}
 \partial_t{\cal P}=\beta\partial^2_{\chih}{\cal P} + \frac{\alpha}{2}\partial_{\chih}[V'{\cal
 P}].
\end{equation}

It follows from the Langevin equation (\ref{KGsmooth2}) with a
top-hat window function for which the noise is given by
Eq.~(\ref{noise4}) with $\beta=-1/3H$ and $\alpha=H^{3/2}/2\pi$
that the one-point PDF, ${\cal P}(\chih,t)$, is solution of
\begin{equation}\label{fp}
 \partial_t{\cal P}=\frac{H^3}{8\pi^2}\partial_{\chih}^2{\cal P}+
 \frac{1}{3H}\partial_{\chih}\left(V'{\cal P}\right).
\end{equation}
In a cosmological context, this equation was first derived in
Ref.~\cite{staro0} in the case $H={\rm const.}$, which we are
interested in, and then in Ref.~\cite{vilenkin} in the case
$V'=0$. It was generalized to more involved situations in
Ref.~\cite{linde}.

The solution of Eq.~(\ref{fp}) where studied in Ref.~\cite{staro2}
in which it is shown that $\mP$ approaches the static equilibrium
solution
\begin{equation}\label{peq}
 {\cal P}_{_{\rm eq}}={\cal N}^{-1}\exp\left(-\frac{8\pi^2}{3H^4}V(\chih)\right),
 \qquad
 {\cal N}\equiv\int_{-\infty}^\infty
 \exp\left(-\frac{8\pi^2}{3H^4}V(\chih)\right)\dd\chih
\end{equation}
irrespectively of the initial conditions.

For the quartic potential, Eq. (\ref{Vquartic}), we find
\begin{equation}\label{peq2}
 {\cal P}_{_{\rm eq}}(\chih)=\frac{1}{2\Gamma(5/4)H}\left(\frac{\pi^2\lambda}{9}\right)^{1/4}
 \exp\left[-\frac{\pi^2\lambda}{9H^4}\chih^4\right].
\end{equation}

They are few lessons to learn from this result. The excursion
values of $\chih$ are bounded. Their distribution does not depend
on the remote past history of $\chi$ and, importantly for the
following, the typical value one can expect for $\chih$, and
therefore $\chib$, is $H/\lambda^{1/4}$.

%%------------------------------------------------------------
\subsection{Consequences on the shape of the PDF of $\delta\chis$}\label{inf_pdf}

We can then insights into the shape of the probability
distribution function of $\delta\chis$ as a function of $\chib$.

Quantitative results can be drawn from the perturbation approach
we initially developed in our previous work~\cite{bu1}. We expand
the filtered field in terms of the coupling constant as
\begin{equation}\label{chiexp}
  \chis(\eta)=\chis^{(0)}(\eta)+\chis^{(1)}(\eta)+\dots.
\end{equation}
$\chis^{(0)}$ represents the value of the filtered field when the
interaction term is switched off. In the slow-roll regime the
field $\chis$ follows the Klein-Gordon equation~(\ref{KGsmooth2}).
Contrary of the previously studied case of $\chih$, the evolution
equation does not contain any noise term because the smoothing
scale $R$ is now fixed and there is no new modes entering $\chis$.
The free filtered field, $\chis^{(0)}$, is therefore constant and,
from the discussion of the previous section, it cannot be assumed
to be Gaussian distributed with a zero mean: its expectation
value, $\chisOb$, is actually the value of $\chib$ at time
$\eta=1/\rh$. Since this field value is going to be only weakly
affected but its subsequent evolution, in the following we will
identify $\chisOb$ and $\chib$. The difference
$\dchisO\equiv\chis^{(0)}-\chisOb$ has been built up from the
modes that have left the horizon between $-1/\kh$ and $-1/\ks$. It
can be assumed to be Gaussian distributed with a width precisely
given by $\sigma_{\delta}$.

The first order term in $\lambda$, ${\chis}^{(1)}$, evolves
according to
\begin{equation}\label{dsl1evol}
  3H\dot{\chis}^{(1)}=
  -\frac{\lambda}{3!}\left[\chis^{(0)}\right]^3.
\end{equation}
In this approach the treatment of the filtering of the r.h.s. of
this equation is very crude. The results we are going to find can
anyway be checked against more rigorous calculations based on the
computed shape of the tri-spectrum. Our goal now is to capture the
essential effects of a non-vanishing $\chib$ on the statistical
properties of $\dchis$.

The equation of evolution~(\ref{dsl1evol}) can be solved to get
\begin{equation}\label{dsl1sol}
  \chis^{(1)}(t)=-\lambda(t-t_{_{\rm H}})\frac{\left(\chis^{(0)}\right)^3}{18\,H},
\end{equation}
which also reads
\begin{equation}\label{dsl1sol2}
  \chis^{(1)}(t)=-\frac{\lambda N_e}{18\,H^2}\left(\chis^{(0)}\right)^3,
\end{equation}
$N_e$ being the number of $e$-fold between $t_H$ and the end of
inflation.

These results imply that
\begin{equation}\label{chisbar}
  \chib\approx \chisOb-\frac{\lambda N_e}{18\,H^2}\left[\left(\chisOb\right)^3+
  3\,\chisOb\,\sigma_\delta^2
  \right]
\end{equation}
which explicitly shows that $\chib$ and $\chisOb$ are equal at
leading order in $\lambda$. It also gives
\begin{equation}\label{dchis}
  \dchis=\dchisO-\frac{\lambda N_e}{18\,H^2}\left\lbrace
  \left(\dchisO\right)^3+
  3\left[\left(\dchisO\right]^2-\sigma_{\delta}^2\right)\chib +
  3\dchisO\chib^2 \right\rbrace.
\end{equation}
It is straightforward to see that $\dchis$ has acquired a non zero
third order moment,
\begin{equation}\label{skewnessdchis}
  \overline{\dchis^3}=-\frac{\lambda
  N_e}{H^2}\chib\sigma_{\delta}^4
\end{equation}
at leading order in $\lambda$. This is a finite volume effect in
the sense that it exists for a fixed (not ensemble averaged) value
of $\chib$. This effect cannot a priori be neglected. From the
study of the previous paragraph, we know that $\chib$ should be of
the order of $H/\lambda^{1/4}$, which implies  that the reduced
skewness of $\dchis$,
$\overline\dchis^3/(\overline\dchis^2)^{3/2}\sim
\lambda^{3/4}N_e\sigma_{\delta}/H$ is significant as soon as
$\lambda^{3/4}N_e$ approaches unity, a condition similar to that
encountered in Sect.~\ref{subsec_low}.

Actually the evolution equation for $\chis$ can be solved in
\begin{equation}\label{chissol}
\chis=\frac{\chisO}{\sqrt{1-\frac{\lambda\,N_e}{9H^2}\left(\chisO\right)^2}}
\end{equation}
and the distribution of $\chis$ can then be infered from that of
$\chisO$ assuming the latter is Gaussian distributed with a
non-zero mean value~\footnote{As noted in Ref.~\cite{bu1}, such a
simple variable change implicitly incorporates ``loop order"
effects that, because of sub-Hubble physics, are not necessarily
correctly estimated. In that paper we developed a more elaborated
method which allows the reconstruction of the PDF from the only
tree order contributions of each cumulant. As the two approaches
eventually give the same qualitative results we restrict here our
analysis to the simplest method.}.

\begin{figure}
 \centerline{\epsfig{figure=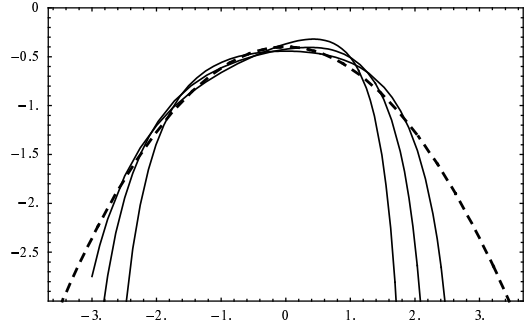,width=7cm}}
 \caption{PDF of $\delta\chis=\delta\chis-\chib$ for different
values of $\chib$. The dashed line corresponds to a Gaussian
distribution; the dot-dashed line to the deformed distribution of
$\dchis$ when $\lambda N_e/H^2=1$ and $\chib=0$ and the solid
lines when the latter equals $0.5$ and $1$.} \label{fvepdf}
\end{figure}

In figure~\ref{fvepdf}, we present the deformation of the PDF of
$\delta\chis$ while $\chib$ is varied. As expected, it shows that
when $\chib$ is not zero, the PDF gets skewed in a way that can be
easily understood: when $\chib$ is positive it gets more difficult
to have excursion towards larger value of $\chis$, but easier to
roll down to smaller values. It is as if the field $\chi$ was
actually evolving in the potential $\lambda(\chi+\chib)/4!$. As a
result one naturally expects the field $\chi$ to have
non-vanishing three-point function. As mentioned before, these
calculations treat the smoothing in a rather simplified way but we
think that for illustrative purposes it encapsulates the main
effects that we want to describe.

It nonetheless shows the way for the computation of the finite
volume effects for expected stochastic properties of the field.

\section{Finite volume effects on high order cumulants}
\label{sec_finitevol}

What the Langevin picture suggests is that finite volume effects
on the observed quantities are not due to the whole stochastic
process that give birth the observed field but mainly to the value
of $\chib$ alone. In other words, we expect to have
\begin{equation}\label{2bis}
 {\cal P}(X_n)=\int {\cal P}(\chib)\,\delta\left(X_n
 -\left<\left(\chis-\chib\right)^n\right>_{\chib}\right)
 \,\dd\chib.
\end{equation}
This form (\ref{2bis}) implies in particular that the PDF of
$X_n$, ${\cal P}(X_n)$, is expected to be peaked around the
expectation value of $\delta\chis^n$ at $\chib$ fixed. In
particular, it implies that
\begin{equation}\label{2ter}
\left< X_n^p\right> \simeq \left<
\left(\left<\delta\chis^n\right>_{\chib}\right)^p\right>
\end{equation}
We will explicitly check such a property for the lower order
cumulants.

In general, for small enough values of $\chib$, the constrained
ensemble averages of the form $\mg A(\chi)\md_{\chib}$ should be
given by
\begin{equation}\label{Aconstrained}
\mg A(\chi)\md_{\chib}\simeq\mg
A(\chi)\chib\md\,\frac{\chib}{\sigma^2_{\chib}}
\end{equation}
where $\sigma^2_{\chib}$ is the variance square of the $\chib$
fluctuations. This relation, exact for Gaussian fields, is only
approximate in general. It can be derived for stochastic variables
following a quasi-Gaussian distribution. Here, it will be valid
only if the excursion values of $\chib$ are modest compared to the
fluctuations of $\dchis$.

Not surprisingly, it implies that the even order cumulants are
left unchanged.

%%------------------------------------------------------------
\subsection{The bi-spectrum}\label{bispec}

Non trivial finite volume effects are then going to appear at the
level of the third order cumulants or correlation functions. In
particular it induces a non-vanishing bispectrum, three-point
correlation function of the wave-vectors, which is going to be
given by
\begin{equation}\label{bispec1}
  \mg\chi(\bk_1)\chi(\bk_2)\chi(\bk_3)\md_{\chib}\simeq
\mg\chi(\bk_1)\chi(\bk_2)\chi(\bk_3)\chib\md_c\
\frac{\chib}{\sigma_{\chib}^2}
\end{equation}
when $\chib$ is small enough. Using Eq.~(\ref{chi4pt}) to express
$\left<\chi_{\bk_1}\ldots\chi_{\bk_4}\right>_c$ and using that
Eq.~(\ref{defchib}) implies that $\chib_\bk=\chi_\bk\widehat
W(k\rh)$, the previous expression reduces to
\begin{eqnarray}\label{bispec2}
\mg\chi(\bk_1)\chi(\bk_2)\chi(\bk_3)\md_{\chib}&=&
   -\frac{\lambda}{3H^2}(2\pi)^{9/2}\, \frac{\chib}{\sigma_{\chib}^2} \,
 \log\left[\left(\sum_i k_i+\left\vert\sum_i\bk_i\right\vert\right)\eta\right]
 \times\nonumber\\
 &&
\left\lbrace
   P(k_1)P(k_2)P(k_3) +
   P\left(\left\vert\sum_i \bk_i\right\vert\right)\left[P(k_1)P(k_2)+
  \hbox{perm.}\right]\right\rbrace\widehat
  W\left(\left|\sum_i\bk_i\right|\rh\right).
\end{eqnarray}
The factor $\widehat
W\left(\vert\bk_1+\ldots\bk_3\vert\rh\right)$, that arises from
the contribution of modes with $k<1/\rh$ to $\chib$, ensures that
$\vert\bk_1+\ldots\bk_3\vert$ is small compared to each of the
$k_i$, and thus can be neglected in the log term. It implies that
for an Harrison-Zel'dovich type spectrum that
$P(\vert\bk_1+\ldots+\bk_3\vert)\gg P(k_i)$ so that the first term
of (\ref{bispec2}) is negligible. As a result we deduce that
\begin{equation}\label{bispec3}
\mg\chi(\bk_1)\chi(\bk_2)\chi(\bk_3)\md_{\chib}\simeq-\frac{\lambda
\chib}{3 H^2}(2\pi)^{9/2}\log\left[\left(\sum_i
k_i\right)\eta\right]\left[P(k_1)P(k_2)+\hbox{perm.}\right]
\frac{P\left(\left\vert\sum_i \bk_i\right\vert\right) \widehat
W\left(\left\vert\sum_i
\bk_i\right\vert\rh\right)}{\sigma_{\chib}^2}.
\end{equation}
Now, noting that, by definition, $P(k)\widehat
W(k\rh)\dd^3\bk/\sigma_{\chib}^2$ integrates to unity and that the
function ${P\left(\left\vert\sum_i\bk_i\right\vert\right) \widehat
W\left(\left\vert\sum_i\bk_i\right\vert\rh\right)}$ is, for the
modes we are interested in, peaked near the origin we obtain that
this factor is essentially equal to
$\delta\left(\sum_i\bk_i\right)$. It therefore implies
that
\begin{equation}\label{bispec4}
\mg\chi(\bk_1)\chi(\bk_2)\chi(\bk_3)\md_{\chib}=-\frac{\lambda
\chib}{3 H^2}(2\pi)^{9/2}\,\log\left[\left(\sum_i
k_i\right)\eta\right]\,\left[P(k_1)P(k_2)+\hbox{perm.}\right]
\,\delta\left(\sum_i\bk_i\right).
\end{equation}
Here is one of the main points of this paper: Finite volume
effects induce a non-vanishing three-point function although the
potential in which $\chi$ evolves is symmetric.

From this bispectrum it is possible to compute the third order
moment of $\dchis$. Its amplitude will be in agreement with what
had been obtained from the Langevin equation. It is also the
three-point function one expects for a field evolving in potential
$\lambda\,\chib\,\chi/3!$ (see Ref.~\cite{bbu} for details).

\subsection{The three-point cumulant}\label{nu3}

We now turn to the lowest order cumulant exhibiting a non-trivial
result due to a non-vanishing $\chib$, that is $X_3$, which reads
\begin{equation}
 X_3 = \int\frac{\dd^3\bk_1\ldots\dd^3\bk_3}{(2\pi)^{9/2}}
 \widehat W(|\bk_1+\bk_2+\bk_3|\rh)
 \widetilde W(k_1)\widetilde W(k_2)\widetilde W(k_3)\chi_{\bk_1}\chi_{\bk_2}\chi_{\bk_3}.
\end{equation}
Obviously its ensemble average vanishes,
$$
\left<X_3\right> = 0
$$
but not $\left<X_3^2\right>_c$. Let us check, as expected from our
analysis [see Eqs.~(\ref{2bis}-\ref{2ter})], that it is well
approximated by
$\left<\left(\left<X_3\right>_{\chib}\right)^2\right>$.

To evaluate the latter expression, we start from the expression of
$\left<X_3\right>_{\chib}$  that is defined by
\begin{equation}
\mg X_3\md_{\chib} \simeq\frac{\chib}{\sigma_{\chib}^2}
\int\frac{\dd^3\bk_1\ldots\dd^3\bk_3}{(2\pi)^{9/2}}
 \widehat W(|\bk_1+\ldots+\bk_3|\rh)
 \widetilde W(k_1)\ldots\widetilde W(k_3)
 \mg\chi_{\bk_1}\ldots\chi_{\bk_3}\md_{\chib}.
\end{equation}
Using Eq.~(\ref{bispec4}), it reduces after integration over $\bk_3$ to
\begin{equation}
\mg X_3\md_{\chib} \simeq -\frac{\lambda\chib}{3H^2}
\int\dd^3\bk_1\dd^3\bk_2
 \widetilde W(k_1)\widetilde W(k_2)\widetilde W(|\bk_1+\bk_2|)
 \left\lbrace P(k_1)P(k_2) + P(|\bk_1+\bk_2|)\left[P(k_1) + P(k_2)
 \right] \right\rbrace
\end{equation}
where the window function $\widehat W(|\bk_1+\ldots+\bk_3|\rh)$ was
aborbed during the integration over $\bk_3$ due to the Dirac distribution.
Such an expression can be computed following the same lines as for
the computation of the expectation value of the fourth order
cumulant (\ref{kurtosisexpression}),
\begin{equation}
\mg X_3\md_{\chib} \simeq -{\lambda \chib\over
3H^2}\left(\frac{H^2}{2}\right)^2 \int_{\kh}^{\ks}{\dd^3
\bk_1\over k_1^3} \int_{\kh}^{\ks}{\dd^3 \bk_2\over k_2^3}\
3\mC\left(\vert\bk_1+\bk_2\vert\right)\
\log\left[\left(k_1+k_2+\vert\bk_1+\bk_2\vert\right)\eta\right].
\end{equation}
where, again, $\mC\left(\vert\bk_1+\bk_2\vert\right)$ simply expresses the
conditions, $\kh < \vert\bk_1+\bk_2\vert < \ks$. A simple
expression for this integral can be obtained when $\ks$ is much
larger than $\kh$ where it is possible to replace $\bk_1+\bk_2$
and $k_1+k_2+\vert \bk_1+\bk_2\vert$ by respectively either
$\bk_1$ and $2\,k_1$ or $\bk_2$ and $2\,k_2$. Finally the integral
reads,
\begin{equation}\label{skewnessexpression}
\mg X_3\md_{\chib}=-\lambda\, \chib
(4\pi)^2\left(\frac{H^2}{2}\right)^2
    \log^2\left({\ks\over\kh}\right)
    \log\left(2\eta\ks^{2/3}\kh^{1/3}\right)
\end{equation}
e.g.
\begin{equation}\label{skewnessexpression2}
\mg X_3\md_{\chib}= - {\lambda\ \chib\ \log\left(2\eta
\ks^{2/3}\kh^{1/3}\right)}\frac{\sigma_{\delta}^4}{H^2}
\end{equation}
which reproduces the result (\ref{skewnessdchis}) if $N_e$ is
identified with $\log\left(2\eta \ks^{2/3}\kh^{1/3}\right)$. In conclusion,
we end up with
\begin{equation}\label{TT}
\left<\left(\left<X_3\right>_{\chib}\right)^2\right>^{1/2}\simeq
 {\lambda\  \log\left(2\eta \ks^{2/3}\kh^{1/3}\right)}
 \frac{\sigma_{\delta}^4\sigma_{\chib}}{H^2}
\end{equation}

In this case, it is actually possible to compute
$\left<X_3^2\right>$ from a perturbation theory approach in order
to check that its dominant contribution is indeed
$\left<\left(\left<X_3\right>_{\chib}\right)^2\right>$.
$\left<X_3^2\right>$ is given in general by
\begin{eqnarray}
 \left<X_3^2\right> &=& \int\frac{\dd^3\bk_1\ldots\dd^3\bk_3}{(2\pi)^{9/2}}
 \int\frac{\dd^3\bk'_1\ldots\dd^3\bk'_3}{(2\pi)^{9/2}}
 \widetilde W(k_1)\ldots\widetilde W(k_3)\widetilde W(k'_1)\ldots\widetilde W(k'_3)\nonumber\\
 &&\qquad\widehat W(|\bk_1+\ldots+\bk_3|\rh)\widehat W(|\bk'_1+\ldots+\bk'_3|\rh)
 \left<\chi_{\bk_1}\ldots\chi_{\bk_3}\chi_{\bk'_1}\ldots\chi_{\bk'_3}\right>_c.
\end{eqnarray}
It involves the expression of the 6-point correlation function
$\left<\chi_{\bk_1}\ldots\chi_{\bk_3}\chi_{\bk'_1}\ldots\chi_{\bk'_3}\right>_c$
which, in a perturbation theory approach, can be split into 2
contributions
$$
 I\equiv\left<\chi^{(0)}_{\bk_1}\chi^{(0)}_{\bk_2}\chi^{(1)}_{\bk_3}\chi^{(0)}_{\bk'_1}
 \chi^{(0)}_{\bk'_2}\chi^{(1)}_{\bk'_3}\right>,\qquad
 II\equiv\left<\chi^{(0)}_{\bk_1}\chi^{(0)}_{\bk_2}\chi^{(2)}_{\bk_3}\chi^{(0)}_{\bk'_1}
 \chi^{(0)}_{\bk'_2}\chi^{(0)}_{\bk'_3}\right>
$$
where
$$
 \chi^{(1)}(\bx)\sim-\frac{\lambda}{18}\frac{N_e}{H^2}\left[\chi^{(0)}(\bx)\right]^3,\qquad
 \chi^{(2)}(\bx)\sim \frac{\lambda^2}{8}\frac{N_e^2}{H^4}\left[\chi^{(0)}(\bx)\right]^5.
$$
This implies that
\begin{equation}
 I = (2\pi)^9\frac{\lambda^2}{18^2}18^2\frac{N_e^2}{H^4}
      P(k_1)P(k_2)P(\bk_3+\bk_1+\bk_2|)P(k_2')P(k'_3)\,
      \delta\left(\sum\bk_i + \sum\bk'_i\right).
\end{equation}
and
\begin{equation}
 II = (2\pi)^{9} \frac{\lambda^2}{8}\,6!\,\frac{N_e^2}{H^4}
      P(k_1)P(k_2)P(k_1')P(k_2')P(k'_3)\,
      \delta\left(\sum\bk_i + \sum\bk'_i\right)
\end{equation}
where $18^2$ and $6!$ are symmetry factors.

Let us evaluate the first contribution
\begin{eqnarray}
 \left<X_3^2\right>_c^{(I)} &=&  \lambda^2 \frac{N_e^2}{H^4}
 \int \dd^3\bk_1 \dd^3\bk_2
      P(k_1)P(k_2)\widetilde W(k_1)\widetilde W(k_2)
 \int \dd^3\bk_3
      P(|\bk_3+\bk_1+\bk_2|)\widetilde W(k_3)\widehat W^2(|\bk_1+\bk_2+\bk_3|\rh)
 \nonumber\\
&&
 \int \dd^3\bk'_1 \dd^3\bk'_2
     P(k'_1)P(k'_2)\widetilde W(k'_1)\widetilde W(k'_2)
     \widetilde W(|\bk_1+\bk_2+\bk_3+\bk'_1+\bk'_2|.
\end{eqnarray}
The term $\widehat W^2(|\bk_1+\bk_2+\bk_3|\rh)$ implies that $|\bk_1+\bk_2+\bk_3|\ll\rh$
so that $\widetilde W(k_3)\sim\widetilde W(|\bk_1+\bk_2|)$ and
$\widetilde W(|\bk_1+\bk_2+\bk_3+\bk'_1+\bk'_2|\sim\widetilde W(|\bk'_1+\bk'_2|)$.
Setting ${\bf e} =  \bk_1+\bk_2+\bk_3$, we conclude that
\begin{eqnarray}
 \left<X_3^2\right>_c^{(I)} &\sim&\lambda^2 \frac{N_e^2}{H^4}
      \left[\int\dd^3\bk_1 \dd^3\bk_2
      P(k_1)P(k_2)\widetilde W(k_1)\widetilde W(k_2)\widetilde W(|\bk_1+\bk_2|) \right]^2
      \,
  \int\dd^3{\bf e} P(e) \widehat W^2(e\rh).
\end{eqnarray}
The integral over ${\bf e}$ reduces to $\sigma_{\chib}^2$ so that
\begin{eqnarray}
 \left<X_3^2\right>_c^{(I)} &\sim&
  \lambda^2 \frac{N_e^2}{H^4}\,\sigma_{\chib}^2
   \left[\int \dd^3\bk_1\dd^3\bk_2
      P(k_1)P(k_2)\widetilde W(k_1)\widetilde W(k_2)\widetilde W(|\bk_1+\bk_2|) \right]^2.
      \label{X3squareI}
\end{eqnarray}

The second contribution reduces to
\begin{eqnarray}
 \left<X_3^2\right>_c^{(II)} &\sim&90\lambda^2\frac{N_e^2}{H^4}
 \int \dd^3\bk_1 \dd^3\bk_2
  P(k_1)P(k_2)\widetilde W(k_1)\widetilde W(k_2)\widetilde W(|\bk_1+\bk_2|)\nonumber\\
 &&
 \int \dd^3\bk'_1 \dd^3\bk'_2\dd^3\bk'_3
 P(k_1')P(k_2')P(k_3') \widetilde W(k_1')\widetilde W(k_2')\widetilde W(k_3')
 \widehat W^2(|\bk'_1+\bk'_2+\bk'_3|\rh).
\end{eqnarray}
The second integral reduces to
$$
\int \dd^3\bk'_1 \dd^3\bk'_2\dd^3
 P(k_1')P(k_2')P(|\bk'_1+\bk'_2|) \widetilde W(k_1')\widetilde W(k_2')\widetilde W(|\bk'_1+\bk'_2|)
 \int\dd^3{\bf e} \widehat W^2(e\rh).
$$
The integral over ${\bf e}$ gives $\sim\rh^{-3}$ so that
\begin{eqnarray}
 \left<X_3^2\right>_c^{(II)} &\sim&90\lambda^2\frac{N_e^2}{H^4}\rh^{-3}
 \int \dd^3\bk_1 \dd^3\bk_2
  P(k_1)P(k_2)\widetilde W(k_1)\widetilde W(k_2)\widetilde W(|\bk_1+\bk_2|)\nonumber\\
 &&
 \int \dd^3\bk'_1 \dd^3\bk'_2
 P(k_1')P(k_2')P(|\bk'_1+\bk'_2|)
 \widetilde W(k_1')\widetilde W(k_2')\widetilde W(|\bk'_1+\bk'_2|).\label{X3squareII}
\end{eqnarray}
Due to the term $\widetilde W(|\bk'_1+\bk'_2|)$, we deduce that,
in the case of a Harrisson-Zel'dovich spectrum,
$(\rs/\rh)^3<|\bk'_1+\bk'_2|^{-3}\rh^{-3}<1$ so that this
contribution is at most equal to the one of Eq.~(\ref{X3squareI}).
As a result we have
\begin{equation}\label{X3squareT}
  \left<X_3^2\right>\simeq\left<X_3^2\right>_c^{(I)}.
\end{equation}
From Eq.~(\ref{X3squareI}), this reduces to
\begin{equation}\label{X3squareT2}
  \left<X_3^2\right>^{1/2}\simeq \lambda\,N_e
  (2\pi)^3 \frac{\sigma_\delta^4\sigma_{\chib}}{H^2}
\end{equation}
that can be identified with the expectation value of
$\left<\delta\chis^3 \right>_{\chib}^2$ over the distribution of
$\chib$, as obtained in Eq.~(\ref{TT}).

This explicit computation shows that, as expected, the
fluctuations of the measured values of $X_3$ are mainly due the
fluctuations of $\chib$. It justifies for instance that one should
expect to see a bispectrum of the form (\ref{bispec4}) for such
inflationary models.

%%%%%%%%%%%%%%%%%%%%%%%%%%%%%%%%%%%%%%%%%%%%%%%%%%%
\section{Conclusions}\label{sec_concl}

In this article we have focused on the phenomenology of the
non-Gaussianity generated in models developed in
Refs.~\cite{bu1,bu2}. Interestingly, whereas the metric
perturbation statistics involve only two microscopic parameters
respectively related to the weight of the non-Gaussian component
and to its PDF, the finite volume effects imply that the
statistical properties of any observational quantity will involve
a third parameter. This new parameter arises from the fact that
the mean value, $\chib$, of the field on the size of the
observable universe does not vanish a priori. Obviously $\chib$
cannot be determined on the basis of any observation. As described
in \S~\ref{sec_langevin}, it implies that the originally symmetric
PDF can be skewed and that this skewness is directly proportional
to $\chib$ that needs to be considered as a new parameter of the
PDF while dealing with observations.

These results open the way for more detailed phenomenological
studies. To see how those properties translate for the temperature
and local density fields is not easy. Such investigation will
probably require numerical tools such that developed
in Ref.~\cite{nonGaussnum}.\\

{\bf Acknowledgements}: We thank Nabila Aghanim and Tristan
Brunier for discussions as well as Alain Riazuelo for early
discussions on the project.

%%%%%%%%%%%%%%%%%%%%%%%%%%%%%%%%%%%%%%%%%%%%%%%%%%%%%%%%%%%%%%%%%%%

\end{document}